\documentclass[prd,preprint,amsmath,nofootinbib,superscriptaddress]{revtex4-1}
\usepackage{graphicx}
\usepackage{bm}
\usepackage{epsfig}

\newcommand{\nn}{\nonumber}

\newcommand{\beq}{\begin{equation}}
\newcommand{\eeq}{\end{equation}}
\newcommand{\bea}{\begin{eqnarray}}
\newcommand{\eea}{\end{eqnarray}}

\begin{document}

\title{Gravitational Anderson  Localization}

\author{Ira Z.~Rothstein\footnote{izr@andrew.cmu.edu}}
\affiliation{Department of Physics, Carnegie Mellon University,
    Pittsburgh, PA 15213}

%%%%%%%%%%%%%%%%%%%%%%%%%%%%%%%%%%%%%%%%%%

\begin{abstract}
We present a  higher dimensional model where  gravity is bound to a brane  due to Anderson localization. 
The extra dimensions are taken to be a disordered crystal of branes, with randomly distributed tensions  of
order  the fundamental scale.  Such geometries bind the graviton  and thus  allow for arbitrarily large extra dimensions even when the curvature is small. 
Thus this model is quite distinct from that of Randall and Sundrum where localization is a consequence
of curvature effects in the bulk.
   The hierarchy problem can be
solved by having the standard model brane live a distance away from the brane on which
the graviton is localized. 
The statistical properties of the system are worked out  and it is shown that the scenario leads to a continuum of four dimensional
theories with  differing strengths of gravitational interactions.  We live on one particular brane whose gravitational constant is $G_N$.
\end{abstract}

\maketitle

\section{Introduction}
Allowing for  extra dimensions introduces new perspectives on the rich structure of gravity.  Kaluza and Klein  (KK) proposed a  picture of a low energy universe that arises from the  compactification
of a higher dimensional space-time. In the ``standard'' KK scenarios, we insure four dimensional phenomenology   
by compactifying the extra dimensions on scales   small
enough that   standard model (SM) KK modes  escape detection due to their large masses.  The most natural size of the extra dimensions
is the d dimensional Planck scale ($M_\star$), which can lead to 
strongly coupled gravity, making it difficult to find explicit ground states, a noted exception being
the Horava-Witten solution \cite{HW}.
More recently \cite{largextraD} it was pointed out  that the size of the extra dimensions
can be arbitrarily large, if we  localize our SM fields onto a  D-brane \cite{pol}, surfaces on which strings end. 
 Large extra dimensional models solve the hierarchy problem by diluting the strength of gravity.  Even if $M_\star\sim TeV$, 
the correct value for $G_N$ is generated if  the size ($r_0$) of the extra dimensions obeys
 $
 G_N^{-1}\sim M_\star^{(d-2)}r_0^{(d-4)}.
 $
 %\beq r_0 \sim \left(10^{38}/(10^3)^{n-2}\right)^{1/(n-4)} GeV^{-1} \eeq.%
Given that gravity should look four dimensional down  to $mm$ scales \cite{bounds}  the number of
extra dimensions must be at least two in these models.  In addition to the graviton zero mode there are also light KK modes which,  if the curvature of the extra dimensions are small,   
have  unsuppressed couplings to the SM brane, leading to deviations from  Newton's law near the
scale of the KK mass.

This  scenario  does not solve the hierarchy problem
until we can understand why the extra dimensions are  large compared to the $TeV$ scale.  
To generate a large extra dimension, one needs  some stabilization mechanism. Moreover, to attain a vanishing four dimensional cosmological constant $(\Lambda_4)$, the brane tension must be balanced by
other curvature sources such as a negative bulk cosmological constant or the curvature due to 
the extra dimensions.
  In standard large extra dimensional models the potential for the  size modulos $(r)$ of the extra dimensions is 
\beq
V(r)=f+\Lambda r^n-n(n-1)\kappa M_\star^{n+2}r^{n-2}.
\eeq
Where $f$ is the brane tension, $\Lambda$ is the bulk cosmological constant and $\kappa$ is the curvature of the compact manifold, 
which equals zero or one   for tori and  $n$-spheres, respectively.
%\footnote{ It is also possible
%choose to generate a negative contribution by taking the branes to have negative tension which is
%sensible if they are taken to lie at orbifold fixed point where there are no  Goldstone translational
%modes. } 
 $\Lambda_4=V(R_{min})\sim0$ is obtained by a tuning.

\begin{figure}[thb]
\centering
\includegraphics[scale=.5]{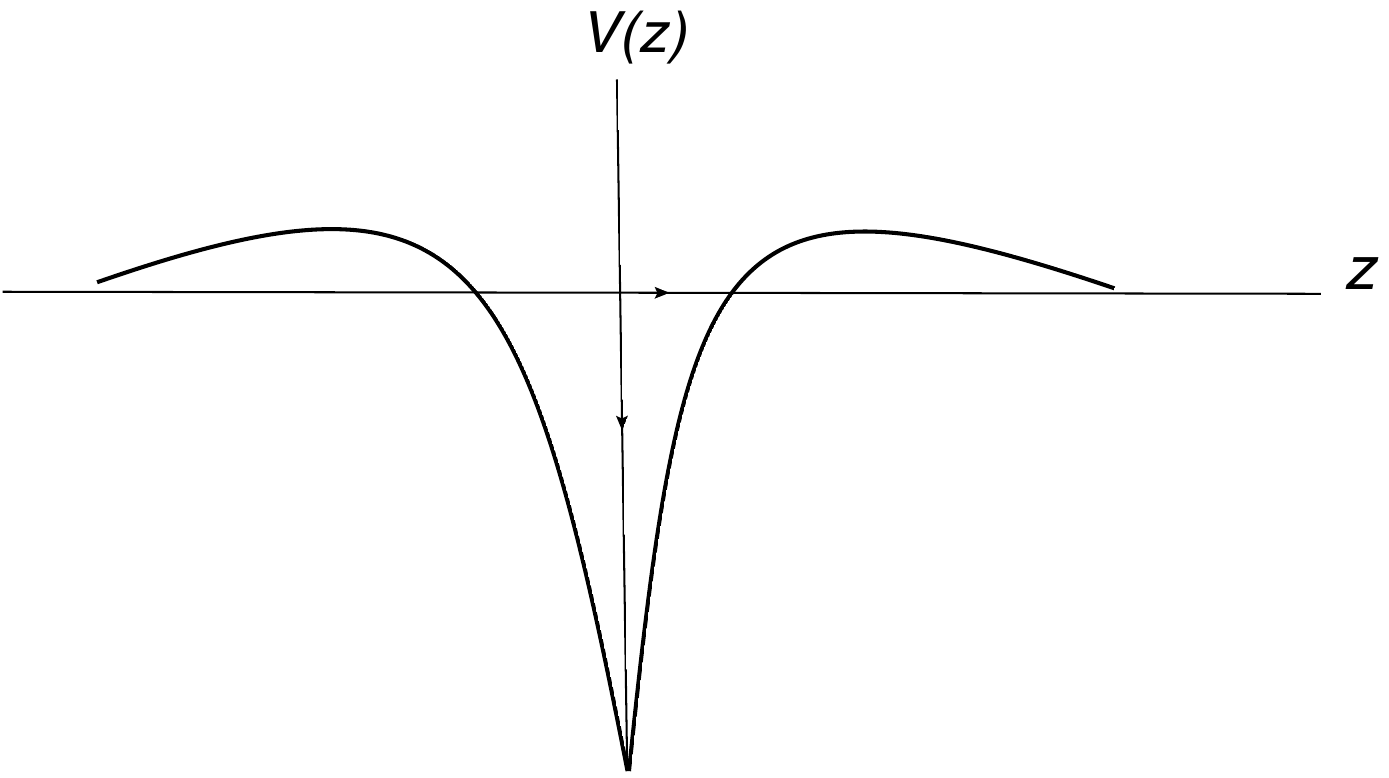}
\caption{The volcano potential generated in the RS scenario which has a zero mode bound state and repels the KK modes.}
\label{volcano}
\end{figure}

In these models  the  bulk is  flat 
%compared to the fundamental scales $M_\star$ in the theory
\footnote{The bulk cosmological constant must thus remain small, which can be accomplished by making the bulk supersymmetric.} and   the graviton is  delocalized, forcing the space
to be compact.
 Randall and Sundrum   (RS)  pointed out \cite{RS1,RS2} that one can make the extra dimensions
arbitrarily large, by utilizing the curvature to localize the graviton, and suppress the contributions of the light KK modes.
%To accomplish this goal one might start with a single brane which, acting as an attractive delta function 
%potential, would have a single (at least in one extra dimension) bound mode. Of course, this won't work since this can not %lead to a stable solution
%to Einsteins' equation and moreover we would not recover Newtons law at long distances.  
%To accomplish this goal  one needs to generate an effective potential which would  a) bind the zero mode
 %b) allow for  a static (or allow for FRW) solution, c) make the effective four dimensional constant vanishingly
%small, d) Suppress the contribution of any KK modes of the graviton that are light.
% (RS) considered the case of a co-dimension one brane where the tension  is balanced by an  anti-DeSitter  bulk    such that 
 %$\Lambda_4 \sim 0$.
  The wave equation for the zero mode maps to the problem of
solving the Schr$\ddot {\rm o}$dinger equation with
a potential with a   ``volcano" shape which binds the zero mode and repels the KK modes (see figure (\ref{volcano})).
%\beq
%V(z)= -\frac{3}{2}k\delta(z) + \frac{15k^2}{(8k \mid \!z\! \mid +1)^2}
%\eeq
%where $z$ is the coordinate in the extra dimension $k$ is the AdS curvature \footnote{ the absolute value arises as a consequence of an imposed $Z_2$ reflection symmetry.}.
In the limit of an arbitrarily large extra
dimension, there is no gap. We recover a sensible low energy theory of
gravity, because the continuum KK modes overlap with the  brane is  suppressed.
%In addition,
% the localized solution allows one to  solve the hierarchy problem by assuming that we live
%on a 3-brane away from the brane upon which the graviton is localized. The  overlap 
%of the graviton  with the SM  fields leads to an   exponentially weaker gravitational force. 
%The hierarchy problem
%is then solved once we employ a ``Goldberger-Wise" \cite{GW} type of mechanism to stabilize the distance between
%the two brane at an natural proper distance.

%%%%%%%%%%%%%%%%%%%%%%%%%%%%%%%%%%%%

We  consider an alternative method of localization that operates even within flat backgrounds.   Linearized
gravitational fluctuations propagate as waves, just like light,  which  can be localized using photonic
crystals \cite{photonicrystals}.
 It is natural 
to ask whether we may  utilize such a mechanism to localize the graviton.
  In photonics crystals    light  is trapped in a  geometry where
excitations in a  given frequency range are  excluded because they  lie in the band gap. In the gravitational case,  this band gap localization will not do, since we are interested in localizing  a ``zero energy'' \footnote{Here the term energy is used only to make an analogy with the quantum mechanics problem.} state at the bottom of a band.

%%%%%%%%%%%%%%%%%%%%%%%%%%%%%%%%%%%%
\begin{figure}[thb]
\centering
\includegraphics[scale=.7]{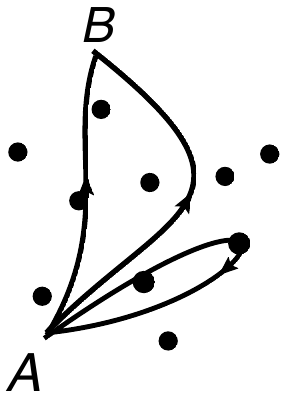}
\caption{The propagation from $A$ to $B$ receives contributions from multiple paths each of which
contributes a random phase to the probability amplitude. Closed paths on the other hand all have time reversed paths
which have identical phases, leading to constructive interference and localization.}
\label{random}
\end{figure}
%%%%%%%%%%%%%%%%%%%%%%%%%%%%%%%%%%%%

To achieve localization we utilize a different approach. Instead of propagating within a uniform crystal we allow for
disorder, which introduces localized states\cite{anderson}. The mechanism for  localization is illuminated in the  propagator where paths which return upon themselves are enhanced relative to other paths, since they receive constructive
interference with time reversed paths  (see fig.\ref{random}).
One can apply a similar reasoning to gravity.  

We begin by noting that   brane-crystals have  been utilized to generate
large stable extra dimensions \cite{stabilization,Corley}. In this scenario one considers a space with $N$ branes
separated by  an amount of order   the fundamental scale $M_\star$. In fact the separation should be slightly
larger then this scale 
to insure that no light string modes  show up in the effective four dimensional low energy theory.
 Inter-brane forces  stabilize the crystal.
  When the graviton is {\it not} localized the
extra dimensions must be compact and thus we cannot stabilize the crystal using charges associated with gauge fields 
due to  Gauss' law.  However,  D-brane charges living in a K-theory group carry charges  which are not associated with a gauge
symmetry\cite{Witten}.  In this case the D-branes will  experience a Van der Waals interaction which, in combination with a hard core repulsive
interaction, can lead to stable lattices   \cite{Corley} . 
In these models Bloch's theorem implies that all the modes are completely delocalized, and as such, 
the space must be compact.
In our model since the   graviton will be localized,  the extra dimensions can be non-compact and
standard (non-BPS) D-brane charges can be utilized.

We now ask whether or not a disordered crystal generates a sensible low energy theory, and, if so,
can it explain the relative weakness of gravity?
In a 
co-dimension one disordered crystal   {\it all} states \cite{Mott} are  localized such that the wave function behaves as
$
\psi(x) \sim \exp(-x/L_{loc}),
$
where $L_{loc}$ is the localization length.  
Note that the Mermin-Wagner-Coleman
theorem will not apply in our case,  since  the fluctuations of the order parameter depend upon   the transverse coordinates.  Furthermore, if the branes are at orbifold fixed points  there is no Goldstone mode to destabilize
long range order.

One might think that such a system is intractable, however, symmetries and simple robust physical arguments  allow us to make quantitative statements.
 We take as our action
\bea
S&=& - \int d^{5}x \sqrt{G}(M_\star^{3} {\cal R})+\sum_{\langle i j\rangle}M_\star^4 V(\mid X_i-X_j\mid)\nn \\
&-& \sum_i \int d^4x\sqrt{ g^{ind}}f_i,
\eea
where $f_i$ are a random distribution of brane tensions which are all assumed to be of order $M_\star$
and $V$ is the above mentioned nearest neighbor  inter-brane potential. 
$G$ is the bulk metric while $g^{ind}$ is the induced metric on the brane.
In general one would expect a bulk cosmological constant, but the robustness of Anderson localization
is such that it will not affect our analysis at the qualitative level.
Furthermore, to emphasize that the localization mechanism presented here is a flat space phenomena
we take the bulk curvature to be much smaller then the fundamental scale.
Four dimensional flatness implies we have one (cosmological constant) fine tuning \footnote{In principle some of the $f's$ could be negative at orbifold fixed points.}.
\beq
\sum_i f_i +\sum_{i,j} M_\star^4 V(\mid X_i-X_j\mid)=0.
\eeq
$\mid X_i- X_j \mid\sim a$ is inter-brane distance  of order $\alpha/M_\star$, 
 and $\alpha\sim 10$ to avoid strong gravity issues.

%To analyze this system we will study quantities averaged over a distribution of tensions. 
%We begin by  writing down the metric  between 
%each pair of D branes as in (\ref{metric}) generalizing to allow for a different $A_i(z)$ in each
%inter-brane segment. 
 In general we will have both ``positional'' disorder where
$X_n-X_{n+1}= a+\Delta_n$  and ``amplitude disorder" where
$f_i=f+\eta_i$.  $\Delta_n,\eta_i$ are random variables drawn from a distribution. Correlations between positional and amplitude disorder can lead to anomalously delocalized states. However, these states
reside in the band centers \cite{review} and would not be relevant in the low energy effective theory. For non-BPS D-branes
 there will be no correlation between these two types of disorder since the positional disorder
is equivalent to a disorder in the D-brane charge which is independent of its tension.
$\eta_i$ and $\Delta_i$ are drawn from a distribution which we will take to be uncorrelated (i.e. white noise) such that
$
\frac{\langle \eta_n \eta_m\rangle}{\eta_n^2} =  \delta_{n,m}~ {\rm and }~\frac{\langle \Delta_n \Delta_m\rangle}{\Delta_n^2} =  \delta_{n,m},
$
where the brackets denote statistical averaging over  distributions  $P_2(\eta_n),P_2(\Delta_n)$, respectively. To avoid fine tunings we will assume 
$
f\sim M_\star, ~{\rm and~} \langle \eta_i\rangle= \langle \Delta_n\rangle=0.
$
In principle we could solve for average values of the inter-brane metric, but  for our purposes
this is unnecessary. The physics is sufficiently robust that we can answer all the relevant questions
 without finding the explicit solution to  Einsteins' equations.  

\begin{figure}[thb]
\centering
\includegraphics[scale=.8]{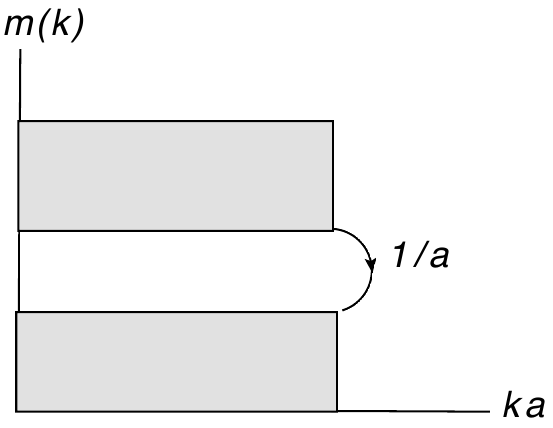}
\caption{The band structure with vanishing disorder. On the scale of $M_\star$ the bands are
nearly continuous. The separation between levels if of order $M_\star/N$. }
\label{spect}
\end{figure}

First we consider  the  ordered (Kronig-Penney like) system.
%The rudimentary relation between the Bloch wave number $k$ and the energy is
%\beq
%\label{disp}
%cos(k a)= cos( m a) + (a f) \frac{sin (m a)}{m a}
%\eeq
%where $f$ is the brane tension, $m$ is the KK mode mass  and $a$ is the lattice spacing. 
The spectrum  (see fig. \ref{spect}), is composed of series of bands and  gaps
both of which 
scale inversely  in the lattice spacing $a$.
%The states within  the bands are separated by an amount of order $M_\star/N$. 
%%%%%%%%%%%%%%%%%%%%%%%%%%%%%%%%%%%%
%%%%%%%%%%%%%%%%%%%%%%%%%%%%%%%%%%%%
 Bloch's theorem implies all states are delocalized, 
thus the extra dimension must be compact 
and  small enough to generate   sufficient mass gap  to reproduce Newton's law down to the $mm$ scale.
In \cite {stabilization,Corley}  the authors solved the hierarchy problem by generating large, but finite,  stable extra dimension 
using $N\sim 10^{32}$ crystal sites in six dimension.  In this model, along with any model which
retains  4-D Poincare invariance, there is  one zero mode.
%The lower band consists of states which are superpositions of the states bound to the delta
%function potentials, while the next band corresponds superpositions of  the set of continuum states.
%Again all of these states are delocalize due to Blochs' theorem.

When we introduce disorder
all  the states are  localized (in 5-D)  which implies that the size of the extra dimension 
no longer controls the strength of gravity.
To calculate the localization length we  first quantify the amount of disorder which is
determined by  the variance ($\sigma$)  of $P(f_i)$.
In our approximation of small tensions we neglect the curvature generated by the branes, in which
case 
this model can be described \cite{review} by  the discrete tightly-bound Anderson model whose 
 Hamiltonian is given by 
\bea
H=\sum_{ij} \epsilon_i a^\dagger_i a_i + t_{ij} a^\dagger_i a^{j+1}.
\eea
$t_{ij}$ is the hopping parameter which is assumed to connect only nearest neighbors (tight binding) and
$\epsilon_i$ are the on-site energies, corresponding to the brane tensions. 
This model, which represents the motion of electrons on a disordered delta function potential lattice,
will properly \footnote{This model does not allow include positional disorder, the effect of
which will not change qualitative results. See for instance \cite{review}.} reproduce our case of interest in the limit  when we can ignore the warping
of the space due to the tensions, i.e.  when $f/m_\star\ll 1$.   Given that we're
assuming that $f$ is small enough that were away from strong coupling, this model
should provide a good qualitative description of spectrum of our theory.

Note that the Anderson model only has one band, but any upper band in our model would play little
to no role in the low energy effective theory.
For our purposes we
take $t_{ij}\approx t \delta_{ij}$ , in which case the system is said to have ``diagonal disorder''.
Furthermore, to avoid any fine-tunings, we will take 
$
P(f_i)
$ to be uniform with  order one variations such that $\langle f_i\rangle=f$.    Note that 
we are assured of one, and only one, zero mode.
More than one massless mode would not lead to a consistent
quantum theory at long distances since there would not be enough ghosts to cancel off all 
of the negative norm states.  The diffeomorphism invariance which assures a sensible low
energy effective theory is however, not manifest in the the Anderson model, which is only
an approximation to our system since we have neglected the curvature induced by the branes.

On dimensional grounds, we would expect the bound
 modes to have masses of order but less then $ M_\star$. Thus the  spectrum is composed of a zero mode
and a gap, though this is just a probabilistic statement. There is a probability to find some low lying
modes that would formally reside in the gap. However, we will see below the density of states in 
the gap is exponentially suppressed.

We are interested in  the ``strong disorder''
limit where
$
\frac{\sigma}{t}\gg 1.
$
The hopping parameter is given by 
\bea
t\equiv \langle i \mid H \mid i+1 \rangle &=& \int_0^a dz \frac{1}{L_{loc}}e^{- (2z+a)/L_{loc}}\nn \\
&=&\frac{a}{2L_{loc}^3}e^{-3a/L_{loc}}(e^{2a/L_{loc}}-1).
\eea
In one dimension transfer matrix techniques allow us to solve for the localization length \cite{Pendry},
$
( L_{loc}/a)^{-1} \sim Log[\sigma a]
$ .
Self-consistency follows since
\beq
\frac{\sigma}{t}= \frac{2 a^4 \sigma^4}{(a^2 \sigma^2-1) 
Log^2(a \sigma)}\sim 2 \frac{(a\sigma)^2}{Log^2(a \sigma)}\sim \alpha^2/Log^2(\alpha)\gg 1.
\eeq
%So we see that we are indeed in the strong disorder regime and the localization length is a few times
%the lattice spaces.

In the disordered case the lower band states are  localized to branes,
while the  upper band consists of would be plane wave modes which are  also localized but randomly
dispersed.
%%%%%%%%%%%%%%%%%%%%%%%%%%%%%%%%%%%%
\begin{figure}[thb]
\centering
\includegraphics[scale=1]{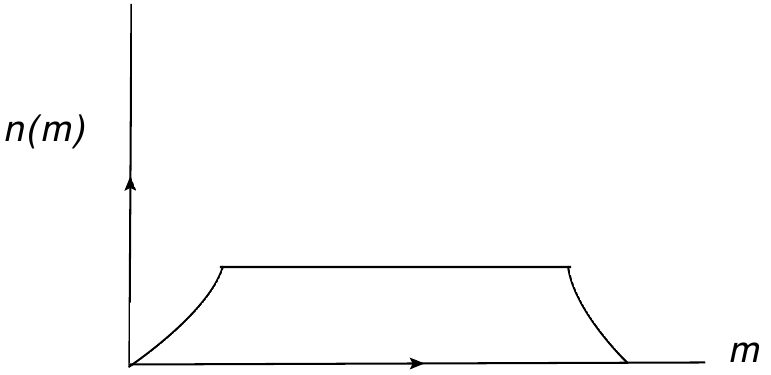}
\caption{The density of states in the strong disorder limit. The width of the flat part of
the band is of order $f$. At the edges the density of states $n(m)$ dies off exponentially
fast and vanishes at the origin. }
\label{denisty}
\end{figure}
%%%%%%%%%%%%%%%%%%%%%%%%%%%%%%%%%%%%

 $G_N$ is fixed by the separation between the brane
which localizes the graviton and the brane upon which the standard model (SM) fields reside.
If the SM lives on a brane which is $p$ lattice spacings away from the gravitons'
brane then the effective four dimensional Planck mass will be
\beq
M_{pl}=M_\star e^{a p/L_{loc}}= M_\star e^{p\log[\alpha]}.
\eeq
Thus to solve the hierarchy problem assuming $M_\star \sim TeV$, the standard model
should live approximately 30 lattice spacings away from the brane on which the zero mode 
graviton is localized.
%We can see that the strength of gravity dies off linearly in distance.
  If the size of the extra dimension is larger
then a $mm$,  we have to be sure that the light KK modes near the band edge do not generate deviations from Newton's
law. In the  non-compact limit the gravitational potential is
\beq
\frac{V_{4}(r)+V_{KK}}{m_1 m_2}= \frac{G_N}{r}+\frac{G_N}{r} \int_{0}^\infty dm  e^{-m r}n(m)\frac{\psi^2_m(0)}{\psi^2_0(0)}
\eeq
where $n(m)$ is the density of states per unit mass, and $\psi_m(0)$ is the wave function  of the mass $m$ KK mode on the  SM brane. 
 The general form  of the Anderson model density of states in the strong disorder limit  is shown in figure (\ref{denisty}) \cite{density}. The density of states is essentially uniform in the region where there masses are of order of
 the brane tensions. It is clear that the contribution from the KK modes can be expected to be negligible
 given that the probability of a very light mode living on a brane sufficiently close to ours that it would
 have a non-negligible overlap is exponentially small. This is particularly true given that the 
 size of the extra dimension can be arbitrarily large.

We hope to have conveyed here is  that there is yet another alternative to compactification beyond the
work of (RS) in which localized gravity can be achieved even in the weak field limit, i.e.
where one can expand around a nearly flat space.
The physics of  the localization is   a consequence of the interference phenomena which
arises due to disorder. The inclusion of curvature is not expected to qualitatively change the physics
as the physical reasoning behind the localization remains the same. Though finding a strongly
curved crystal like solutions to the Einstein equations would be formidable.
 What  is required for the localization is one, or more,
extra dimension  populated by defects. 
As in the RS models, localization allows for arbitrarily large extra dimensions, as opposed to the large extra dimensions scenario \cite{largextraD},    as the gap is not controlled by the size modulos.  The possible difficulties in 
achieving a disordered extra dimension have not been discussed here and deserve further attention.
Also, there are many open phenomenological questions which need to be addressed.
%It is  interesting to note that 
%there are a continuum of worlds in this scenario with a continuum of values for the strength
%of the gravity, we just happen to live on one of them.
\section*{Acknowledgements}
Work supported by \uppercase{DOE}
contracts \uppercase{DOE-ER}-40682-143 and
\uppercase{DEAC02-6CH03000}. The authors gratefully acknowledges helpful conversations with Gil Refael, David Pekker and
Sean Carroll. The authors received  support from the Gordon and Betty Moore foundation and thanks Cal-Tech theory
group for its hospitality.
%%%%%%%%%%%%%%%%%%%%%%%%%%%%%%%%%%%%%%%%%%%%%%%%%%%%%%%%%%%%%
%%%%%%%%%%%%%%%%%%%%%%%%%%%%%%%%%%%%%%%%%%%%%%%%%%%%%%%%%%%%%
%%%%%%%%%%%%%%%%%%%%%%%%%%%%%%%%%%%%%%%%%%%%%%%%%%%%%%%%%%%%%

\end{document}